# Interplay between Superconductivity and Magnetism in $Rb_{0.8}Fe_{1.6}Se_2$ under Pressure.

Vadim Ksenofontov[1], Sergey A. Medvedev*[2,3], Leslie M. Schoop[4], Gerhard Wortmann[5], Taras Palasyuk[2,6], Vladimir Tsurkan[7,8], Joachim Deisenhofer[7], Alois Loidl[7], and Claudia Felser[1,3]

[1]*Institut für Anorganische und Analytische Chemie, Johannes Gutenberg-Universität, D-55099 Mainz, Germany*

[2]*Max-Planck Institute for Chemistry, D-55128 Mainz, Germany*

[3]*Max-Planck Institute for Chemical Physics of Solids, D-01187 Dresden, Germany*

[4]*Graduate School of Excellence Material Science in Mainz, D-55099 Mainz, Germany*

[5]*Department Physik, Universität Paderborn, D-33095 Paderborn, Germany*

[6]*Institute of Physical Chemistry, Polish Academy of Sciences, Warsaw, Poland*

[7]*Experimental Physics V, University of Augsburg, D-86159 Augsburg, Germany*

[8]*Institute of Applied Physics, Academy of Sciences, MD-2028, Chişinău, Republic of Moldova*

**Abstract:**

High-pressure magnetization, structural and $^{57}$Fe Mössbauer studies were performed on superconducting $Rb_{0.8}Fe_{1.6}Se_{2.0}$ with $T_c$ = 32.4 K. The superconducting transition temperature gradually decreases on increasing pressure up to 5.0 GPa followed by a marked step-like suppression of superconductivity near 6 GPa. No structural phase transition in the Fe vacancy-ordered superstructure is observed in synchrotron XRD studies up to 15.6 GPa, while the Mössbauer spectra above 5 GPa reveal the appearance of a new paramagnetic phase and significant changes in the magnetic and electronic properties of the dominant antiferromagnetic phase, coinciding with the disappearance of superconductivity. These findings underline the strong correlation between antiferromagnetic order and superconductivity in phase-separated $A_xFe_{2-x/2}Se_2$ (A = K, Rb, Cs) superconductors.

* s.medvedev@mpic.de

The continuously increasing large family of high-temperature Fe-based superconductors with the highest reported superconducting transition temperature of $T_c$ ~55 K [1] attracts broad scientific interest due to the interplay of superconductivity and magnetism in these compounds. The members of these iron-based superconductors share a common structural motif, namely stacked layers of the $FeX_4$ (X = As or Se) edge-shared tetrahedra, which are considered to be electronically active. The rather simple crystalline structure of the iron-based superconductors favors an understanding of the correlations between their crystalline, magnetic and electronic properties with the ultimate goal to grasp the essentials of the origin of their high-temperature superconductivity.

Recently, new members of Fe-based superconductors, namely $A_xFe_{2-y}Se_2$ (A = K, Rb, Cs and Tl) with $T_c$ values above 30 K, have been found [2]. Neutron studies, μSR spectroscopy, transport, magnetic and calorimetric investigations performed on these systems have shown coexistence between superconductivity and antiferromagnetic (AFM) ordering with relatively high Neel temperatures $T_N$ around 500 K [3]. The coexistence of bulk superconductivity and AFM order with large stable magnetic moments has been put into question by transmission-electron microscopy reporting on a phase separation in the potassium intercalated compound [4]. Subsequent high-resolution nanofocused X-ray diffraction studies provided further experimental evidence that magnetism and superconductivity occur in spatially separated regions [5]. The phase separation scenario is also supported by recent Mössbauer spectroscopy [6] and optical conductivity measurements [7].

One of the outstanding characteristics of iron-based superconductors is a pronounced pressure effect on the superconducting transition temperature. $T_c$ of the simplest Fe-based superconductor, FeSe, amounts 8 K at ambient pressure and reaches 37 K around 8 GPa [8,9]. However, pressure dependent studies of $K_{0.8}Fe_{1.7}Se_2$ [10-12], $Rb_{0.8}Fe_2Se_2$ [13] and $Cs_{0.8}Fe_2Se_2$ [11,14] compounds have shown that in these compounds $T_c$ can only be slightly increased by application of pressure to a maximum value of 33 K and that superconductivity is completely suppressed by further increasing of pressure up to 9 GPa. The origin of the suppression of superconductivity in $A_xFe_{2-y}Se_2$ systems with pressure is still an open question. Here we present the results of combined pressure dependent magnetization, synchrotron X-ray diffraction, and Mössbauer studies of $Rb_{0.8}Fe_{1.6}Se_2$, which clearly indicate that despite the spatial phase separation, superconductivity and AFM order are intimately coupled in these materials.

Single crystals of $Rb_{0.8}Fe_{1.6}Se_2$ were grown by the Bridgman method. Details of preparation and sample characterization were published elsewhere [15]. The single-crystal quality of the grown samples was confirmed by X-ray diffraction. The samples exhibit a transition temperature onset of 32.4 K. For high pressure studies, loading of the high pressure cells was performed in a glove box in an atmosphere of pure nitrogen containing less than 0.1 ppm of oxygen and water to avoid sample decomposition.

Magnetic susceptibility measurements under pressure were performed using a high-pressure cell made from a non-magnetic hardened Cu-Ti alloy equipped with SiC anvils. The diameter of the working surface of the SiC anvils was 0.8 mm, whereas the diameter of the hole in the gasket was 0.3 mm. The cell allows quasi-hydrostatic pressures up to 12 GPa [16]. The hole was filled with a single crystalline $Rb_{0.8}Fe_{1.6}Se_2$ flake and Daphne oil as pressure transmitting medium. The pressure was measured via the Ruby scale from small chips distributed across the sample. The pressure inhomogeneity was determined to be 0.5 GPa across the sample at the highest pressure. $T_c$ was determined from the onset of the superconducting transition curve, i.e. from the intersection of two extrapolated straight lines drawn through the data points in the normal state and through the steepest part in the superconducting state.

High-pressure X-ray diffraction experiments were performed at room temperature on the beamline 01C2 of the NSRRC synchrotron facility, Taiwan. For X-ray diffraction the grained sample of $Rb_{0.8}Fe_{1.6}Se_2$ was loaded in a diamond anvil cell with culets of 450 μm diameter and a tungsten gasket with a sample chamber of 150 μm in diameter. Silicon oil was used as pressure-transmitting medium. The X-ray beam ($\lambda = 0.496$ Å) was collimated to 100 μm, with the image plate detector set perpendicular to the beam. Cerium dioxide was used as external standard to determine the beam center, sample-to-detector distance and tilting angle of the image plate. Collected full-circle powder patterns were processed with FIT2D software.

$^{57}$Fe-Mössbauer spectra were recorded at room temperature using a constant-acceleration spectrometer and a $^{57}$Co(Rh) point source with an active spot diameter of 0.5 mm. Grained $Rb_{0.8}Fe_{1.6}Se_2$ samples were prepared with enriched $^{57}$Fe (20%) and measured in a diamond-anvil pressure cell with silicon oil as pressure-transmitting medium [8]. Due to the granular character of the sample (not finely powdered, but consisting of preferentially oriented single-crystalline flakes, as in the XRD studies), the $^{57}$Fe-spectra exhibit strong texture effects, as described in detail in Ref. 6. These texture effects were carefully taken into account in the spectra analysis. The isomer shift values are quoted relative to those of α-Fe at 295 K.

Figure 1 shows the temperature dependence of the magnetization of $Rb_{0.8}Fe_{1.6}Se_2$ as function of pressure, from which the superconducting transition temperature $T_c$ was derived. Similarly to the observations in the K- and Cs- intercalated superconductors [10-12,14], $T_c$ of $Rb_{0.8}Fe_{1.6}Se_2$ decreases initially quite slowly with increasing pressure. The averaged decrease of $T_c$ up to 5 GPa with a rate of 2.1 K/GPa is followed by a sudden suppression of $T_c$ at pressures near 6 GPa. The pressure dependence of $T_c$ obtained here from the magnetization measurements is in good agreement with recent pressure-dependent electrical resistivity studies of $Rb_{0.8}Fe_2Se_2$ [13], especially with the steep decrease of $T_c$ close to 5 GPa, resembling also the pressure behavior of $T_c$ in the related $Cs_{0.8}Fe_2Se_2$ compound [14]. These observations are in contrast to a more continuous suppression of $T_c$ up to a critical pressure around 9 GPa in isostructural $K_{0.8}Fe_{1.7}Se_2$ [10]. The suppression of superconductivity in $Rb_{0.8}Fe_{1.6}Se_2$ with pressure appears to be irreversible: no superconductivity was observed as the pressure was released from 10.0 GPa to ambient pressure. This is again different to the observations in [10], where after release of pressure $T_c$ reappears.

X-ray diffraction patterns of a grained $Rb_{0.8}Fe_{1.6}Se_2$ sample recorded upon compression indicate the absence of any major structural phase transitions up to pressures of 15 GPa (Fig. 2). Although a rigorous structural refinement cannot be performed due to highly textured sample with different orientations of the flakes hit by the beam at different pressures, the superstructure reflections (110), (020), and (220) corresponding to the *I*4/*m* structure are clearly observed and persist up to the highest pressures indicating the preservation of the vacancy-ordered superstructure up to pressures far above the suppression of superconductivity. This conclusion is well supported by the study of Svitlyk *et al*. where the ordering of the Fe vacancies in $Rb_{0.85}(Fe_{1-y}Se)_2$ has been observed up to ~12.0 GPa [17]. We conclude that the suppression of superconductivity in $Rb_{0.8}Fe_{1.6}Se_2$ is not connected with a structural phase transition in contrast to the structurally related FeSe, in which the transition to the normal conducting state at high pressures is accompanied by a pressure induced structural phase transition [8]. In the present case one must consider other effects connected with changes in the magnetic and electronic properties of the dominant magnetic $\sqrt{5} \times \sqrt{5}$ superstructure responsible for the suppression of superconductivity in the minority phase of $Rb_{0.8}Fe_{1.6}Se_2$. The Mössbauer pressure studies reported below support this suggestion.

Room-temperature Mössbauer spectra recorded at different pressures are shown in Fig. 3. At pressures below 5.2 GPa they consist of a magnetic sextet corresponding to the magnetically ordered component (denoted phase **1**) and a paramagnetic (PM) doublet (denoted phase **2**) with relative fractions of 88(1)% and 12(1)% respectively, as described in Ref. 6. These two components arise due to the phase separation [5,6] during cooling around 550 K. The derived hyperfine parameters for the magnetic hyperfine field $H_{hf}$, the isomer shift IS as well as the quadrupole splitting QS at the lowest pressure of 2.5 GPa are close to those at ambient pressure: $H_{hf}$(**1**) = 252.2(7) kOe, IS(**1**) = 0.53(1) mm/s, QS(**1**) = 1.11(3) mm/s and IS(**2**) = 0.55(2), QS(**2**) = -0.24(2) mm/s.

Remarkable changes in the Mössbauer spectra are observed starting from 5.2 GPa, where an additional new PM doublet appears. The corresponding intensity ratios reveal that this spectral component emerges mostly from the AFM sextet. Therefore this spectral change indicates a magnetic transformation of the AFM phase into a PM state. The hyperfine parameters of this new PM phase (denoted phase **3**) at $p$ = 6.5 GPa are IS(**3**) = 0.50(2) mm/s, QS(**3**) = 0.64(4) mm/s and are very different from those observed in the PM phase **2,** but are close to the parameters of the AFM phase **1** still dominant at this pressure: IS(**1**) = 0.50(1) mm/s, QS(**1**) = 0.85(4) mm/s. This indicates that in the new PM phase **3** the local crystal arrangement of Fe atoms, as well as their electronic properties, namely an $Fe^{2+}$ high-spin state with orbital contributions to the electric field gradient as discussed in Ref. 6, are similar to those in the AFM phase in agreement with the absence of a structural phase transition mentioned above (Fig. 2). The intensity of the new PM fraction **3** progressively increases with increasing pressure and attains 80(1)% of the total spectral area at 13.8 GPa (see Fig. 4a). The transformation of the AFM phase **1** into the PM phase **3** is not complete, 17(1)% of phase **1** can still be observed at $p$ = 13.8 GPa. The fraction of the PM phase **2** decreases similar to phase **1** to 3(1)% at 13.8 GPa. The observed pressure-induced magnetic transition appears to be highly irreversible. The Mössbauer spectrum measured at $p$ = 0.3 GPa after pressure release is dominated by the new PM phase **3** with 53(1)% intensity, while the AFM phase **1** recovers with broadened spectral features to 45(2)% and the component **2** with intensity below 3% can hardly be detected [18].

The pressure dependence of the magnetic hyperfine field $H_{hf}$(**1**) in the AFM phase **1** is presented in Fig. 4(b). While up to 4.2 GPa, the impact of pressure is minor, there is a marked decrease of $H_{hf}$(**1**) = 252(1) kOe at 4.2 GPa to 235(1) kOe at 8.5 GPa, which points to a significant change of the local magnetic and electronic properties at the Fe sites in the √5 x √5 superstructure, the latter reflected also in a concomitant decrease of QS(**1**) = 1.03(4) mm/s at 4.2 GPa to 0.84(4) mm/s at 8.5 GPa. At higher pressures, the variation of $H_{hf}$(**1**) is again very small: $H_{hf}$(**1**) = 236(3) kOe at 13.8 GPa. Of specific interest are the values of the respective hyperfine parameters after release of pressure to 0.3 GPa: $H_{hf}$(**1**) = 240(4) kOe, which is almost identical to the value observed at 13.8 GPa, while the value of QS(**1**) = 1.13(3) mm/s corresponds to the initial ambient pressure value. A similar trend is observed for the new PM phase **3,** where QS(**3**) increased from 0.42(1) mms at 13.8 GPa to 0.77(1) mm/s at 0.3 GPa.

The most prominent feature presented in Fig. 4a,b is the clear relation between the instantaneous disappearance of superconductivity above 5 GPa and the onset of a transformation of AFM phase **1** to a new PM phase **3**, concomitant with a marked reduction of $H_{hf}$(**1**), pointing to a change in the magnetic properties. This fact is especially remarkable in view of the recently established phase separation in alkali-intercalated magnetic superconductors $A_xFe_{2-x/2}Se_2$ [5-7]. According to this concept, only the PM fraction is metallic and superconducting, whereas the major AFM fraction is insulating. Following our recent Mössbauer results [6], the superconducting phase (component **2**) behaves similar to FeSe and therefore it could be expected that a similar scenario of suppression of superconductivity under pressure associated with a structural phase transition will occur [8]. However, as we found in the present investigation, the relatively abrupt and irreversible suppression of superconductivity is associated neither with a structural phase transition in the dominant $\sqrt{5} \times \sqrt{5}$ phase **1,** nor with the disappearance or strong spectral changes of the minority phase **2**. The present data demonstrate that the suppression of superconductivity in $Rb_{0.8}Fe_{1.6}Se_2$ coincides with the onset of a transformation of the dominant AFM phase fraction into a new PM phase and seemingly is connected with change of the local magnetic and electronic properties within the AFM phase. At the present state, we can only propose different reasons for this behavior, obviously closely related to the phase separation with a filamentary superconducting metal embedded in a dominant semiconducting magnetic phase with an ordered $\sqrt{5} \times \sqrt{5}$ superstructure of the Fe vacancies. Due to the irreversibility of the changes induced both on the superconducting and magnetic properties, some irreversible structural changes should be taken into account. Our conclusions can be summarized as follows:

(i) It is remarkable that the magnetic and superconducting properties of this delicate phase mixture stay intact up to relatively high pressures. These properties can be easily changed by pressure induced structural changes within the layers, for instance by diffusion of the Rb ions above 5 GPa. Such a diffusion process of the Rb ions seems to be thermodynamically much more probable than that of Fe ions, as documented by the preservation of the $\sqrt{5} \times \sqrt{5}$ superstructure up to ~15 GPa. Even small changes at the phase boundaries may change the local properties, e.g. a reduction of the Fe moment by bandstructure effects and/or a loss of semiconducting properties. In this context it is interesting to note that a pressure-induced change of semiconducting towards metallic behavior is observed in [10]. To attribute the loss of superconductivity solely to band structure effects by a pressure-induced variation of doping, as proposed in the case of $Rb_{0.8}Fe_2Se_2$ in Ref. 17, is not convincing due to the irreversibility of the loss of superconductivity and magnetic order observed here.

(ii) Another hypothetic reason for the fast disappearance of the superconductivity under pressure could be attributed to the fact that the onset of the magnetic phase transition, seemingly appears at the phase boundaries, eventually connected with metallic properties of the new PM phase **3**, with conduction electrons penetrating the superconducting phase **2**. In this case fluctuations of paramagnetic moments may induce spin flips and hence transfer magnetic fields into the superconducting fraction. Taking into account the high values of the magnetic moment on Fe (ca. 3 $\mu_B$) and corresponding huge exchange/transferred fields far exceeding critical fields, any incomplete compensation of these moments could destroy the neighboring superconducting state [19].

While the suppression of superconductivity in the present $Rb_{0.8}Fe_{1.6}Se_2$ sample is connected with the appearance of the PM phase **3** above 5 GPa, this phase **3** increases to ~80% of the volume of the bulk sample (by further pressure cycling this amount increases to ~90%). This process apparently must be connected with similar changes in the local structure by further Rb diffusion into the former superconducting phase, erasing the local structural, magnetic and electronic differences between the two phases. In this respect, one can suppose that the new PM component **3** observed here at room temperature could be related to the properties of the non-magnetic phase in $AFe_{2-y}Se_2$ systems (A = K, $Tl_{0.6}Rb_{0.4}$) occurring above the suppression of superconductivity and below the reentrant superconductivity at even higher pressures reported in [20]. Therefore investigations of the magnetic and superconducting properties of $Rb_{0.8}Fe_{1.6}Se_2$ at high pressures and low temperatures are presently underway.

Acknowledgement: This work was supported by the DFG within the SPP 1458 by the grants FE 633/10-1, ER 539/6-1 (Mainz) and DE1762/1-1 (Augsburg) and by the TRR80 (Augsburg-Munich). T. P. would like to gratefully acknowledge the support from the Iuventus Plus grant IP2010 030270.

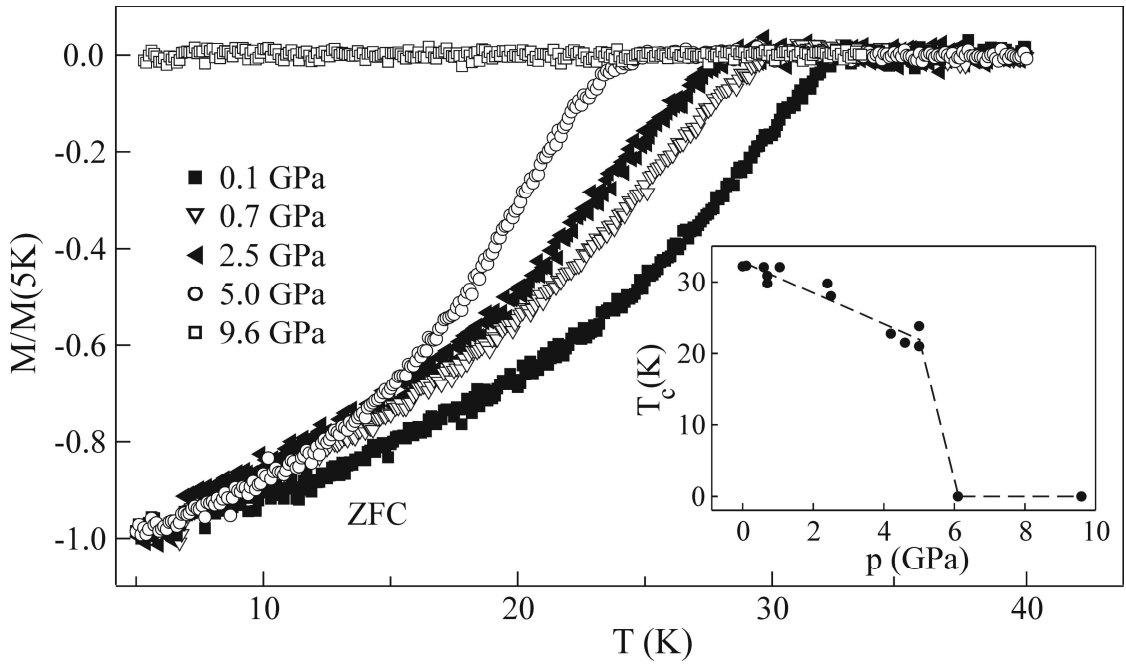

**Fig. 1** Temperature dependence of the magnetization of $Rb_{0.8}Fe_{1.6}Se_2$ at different pressures. ZFC measurements were performed in a magnetic field of 20 Oe, the magnetization was normalized to the values at 5 K. Inset: variation of superconducting transition temperature in $Rb_{0.8}Fe_{1.6}Se_2$ under pressure.

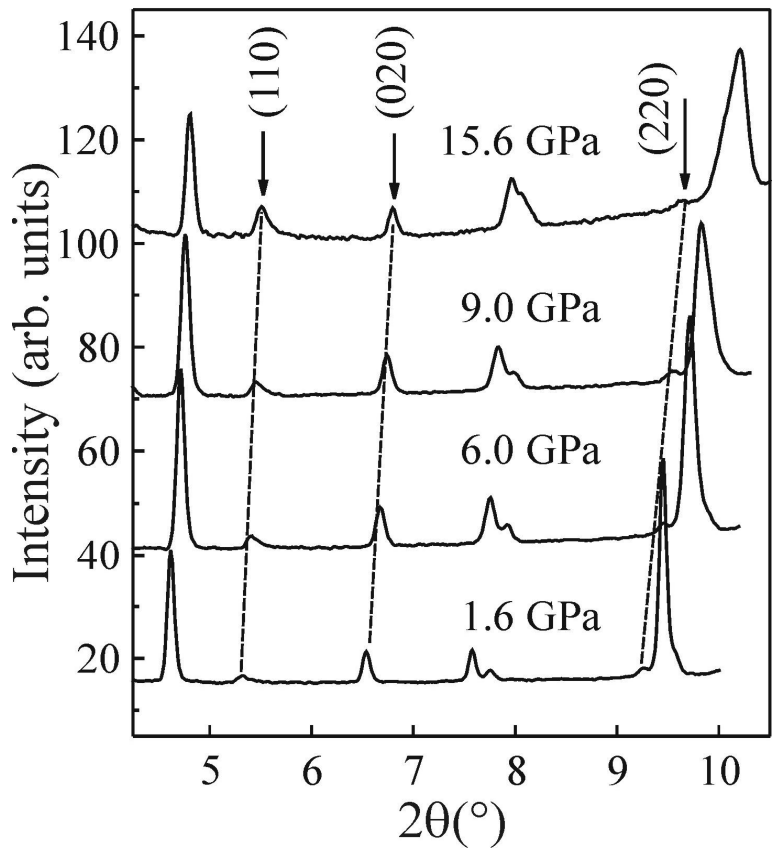

**Fig. 2** Powder diffraction patterns of $Rb_{0.8}Fe_{1.6}Se_2$ at different pressures. Patterns at all pressures can be indexed with *I*4/*m*, corresponding to a $\sqrt{5} \times \sqrt{5}$ vacancy ordered superstructure of the antiferromagnetic phase. The superstructure reflections (110), (020) and (220) indicating a vacancy ordered superstructure, persist up to the highest pressure of 15.6 GPa.

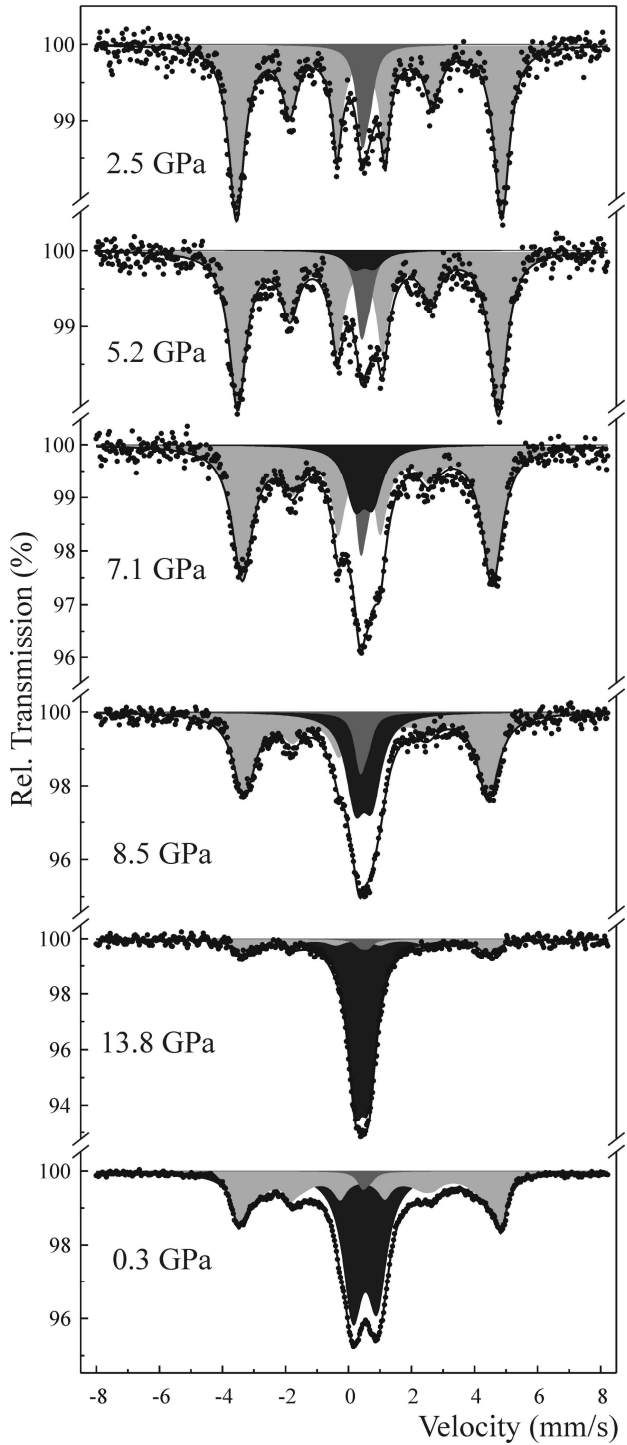

**Fig. 3** Room temperature $^{57}$Fe-Mössbauer spectra of $Rb_{0.8}Fe_{1.6}Se_2$ measured at different pressures. Subspectra of the magnetic Fe sites are marked in light gray (AFM fraction **1**), subspectra of non-magnetic Fe sites (fraction **2**) are shown in gray. At 5.2 GPa a new PM fraction **3** (doublet shown in black) emerges from the AFM fraction **1**. The Mössbauer spectrum measured at $p = 0.3$ GPa after $p = 13.8$ GPa is dominated by the new PM doublet corresponding to the fraction **3**.

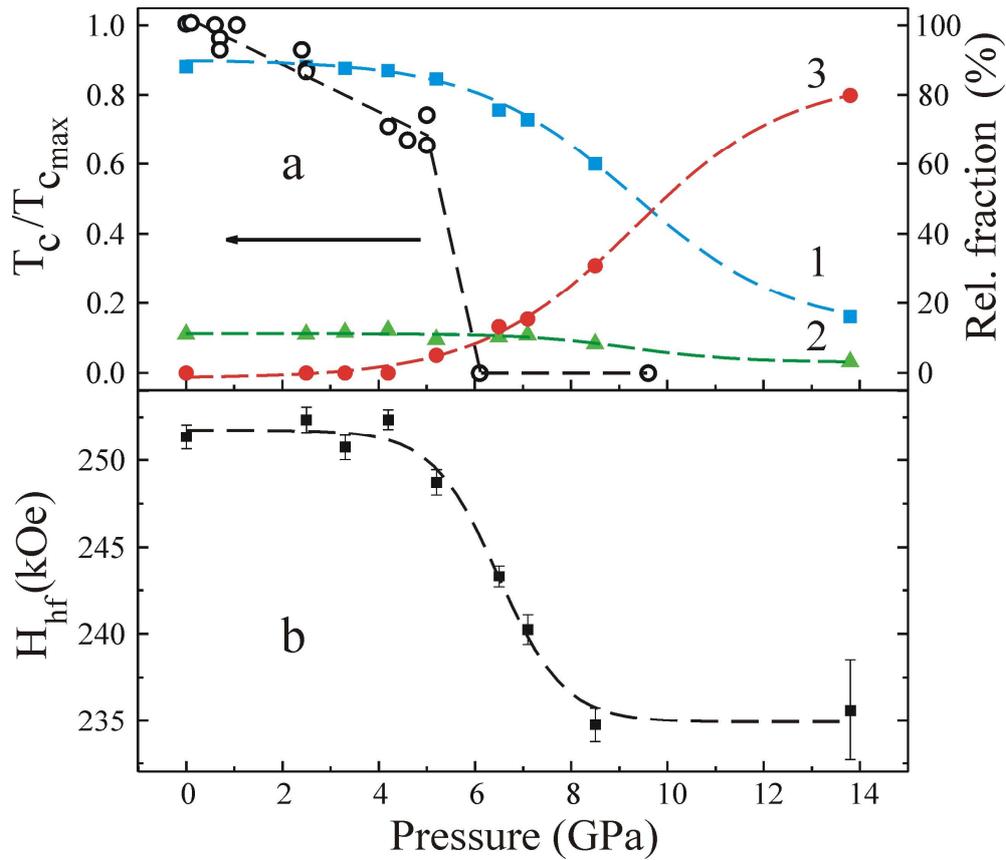

**Fig. 4 (a)** Pressure dependence of PM fraction **2** (triangles, green) together with the PM fraction **3** (closed circles, red), the latter originating mainly from the AFM fraction **1** (squares, blue) plotted together with normalized variation of $T_c/T_{cmax}$ in $Rb_{0.8}Fe_{1.6}Se_2$ (opened circles, black). The onset of the AFM to PM transformation strikingly correlates with the suppression of superconductivity under pressure. **(b)** Pressure effect on $H_{hf}(\mathbf{1})$ of Fe atoms in the AFM phase **1**. The strong decrease of $H_{hf}(\mathbf{1})$ above 5.2 GPa is related to the magnetic transformation. Dashed curves are eye guides.